%% file: fcnc_prl.tex
\documentclass[aps,prl,twocolumn,showpacs,groupedaddress]{revtex4}  
\usepackage{graphicx}  
\usepackage{dcolumn}   
\usepackage{bm}        
\usepackage{amssymb}   
\usepackage{rotating}
\newcommand{\MET} {\mbox{$\not \!\! E_T$}}

\newcommand{\ppbar} {p\bar{p}}
\newcommand{\ttbar} {t\bar{t}}

\def        \dzero  {D0~}

\def        \kappag  {\kappa_g}
\def        \kappac  {\kappa_g^c}
\def        \kappau  {\kappa_g^u}
\def        \kappacLambda  {\kappa_g^c/\Lambda}
\def        \kappauLambda  {\kappa_g^u/\Lambda}

\newcommand{\lsim}{\mathrel{\hbox{\rlap{\lower.55ex\hbox{$\sim$}} \kern-.3em 
\raise.4ex \hbox{$<$}}}}

\begin{document}
%


\hspace{5.2in} \mbox{FERMILAB-PUB-07-031-E}

\title{Search for production of single top quarks via flavor-changing neutral currents 
\\  at the Tevatron}

\input list_of_authors_r2.tex  

\date{February 1, 2007}

\begin{abstract}
We search for the production of single top quarks via flavor-changing
neutral current couplings of a gluon to the top quark and a charm 
($c$) or up ($u$) quark. We analyze  230~pb$^{-1}$ of lepton$+$jets 
data from $\ppbar$ collisions at a center of mass energy of 1.96 TeV 
collected by the D0 detector at the Fermilab Tevatron Collider.  We 
observe no significant deviation from standard model predictions, and 
hence set upper limits on anomalous coupling parameters 
$\kappacLambda$ and $\kappauLambda$, where the $\kappag$ 
define the strength of the $tcg$ and $tug$ couplings, and 
$\Lambda$ defines the scale of new physics. The limits at 
$95\%$ C.L. are: $\kappacLambda < 0.15~\rm TeV^{-1}$ 
and $\kappauLambda < 0.037~\rm TeV^{-1}$. 
\end{abstract}

\maketitle

\clearpage
\normalsize

Top quarks were discovered  in 1995 by the CDF and  \dzero  collaborations~\cite{topdiscovery} at the Fermilab Tevatron Collider in $\ttbar$ pair production involving strong interactions. The standard model (SM) also predicts the production of single top quarks via electroweak exchange of a $W$ boson with cross sections of 0.88 pb in the $s$-channel ($tb$) and 1.98 pb in the $t$-channel ($tqb$)~\cite{sintop-xsecs}. At the $95\%$ C.L., limits set by \dzero are  6.4~pb on the $s$-channel cross section and 5.0~pb on the $t$-channel cross section~\cite{Abazov:2005zz}, and those set by CDF  are 13.6~pb and 10.1~pb, respectively~\cite{RunII:cdf_result}. \dzero recently reported evidence for the production of single top quarks at significance of 3.4 standard deviations~\cite{singletopPRL}.

Since the top quark's discovery, several precision measurements have been made of its properties. Its large mass close to the electroweak symmetry-breaking scale suggests that any anomalous coupling is likely to be observed first in the top quark sector. One form of anomalous couplings can give rise to a single top quark in the final state through flavor-changing neutral current (FCNC) interactions with a charm or an up quark, involving the exchange of a photon, a $Z$ boson, or a gluon~\cite{tait}. Although such interactions can be produced by higher-order radiative corrections in the SM, the effect is too small to be observed~\cite{eilam}. Any observable signal indicating the presence of such couplings would be evidence of physics beyond the SM and would shed additional light on flavor physics in the top quark sector. 

At present, strong constraints exist for FCNC processes via a  photon or a $Z$ boson exchange~\cite{cdfrun1_fcnc,lep_fcnc,hera_fcnc} from studies of both the production and decay of top quarks.  In this Letter, we present a search for production of single top quarks via FCNC couplings of a gluon to the top quark in data collected from  $\ppbar$ collisions at $\sqrt{s} = 1.96~\rm TeV$ using the \dzero detector. This is the first search of its kind at hadron colliders. We consider top quark production rather than decay, since the former is more sensitive to the anomalous couplings ($\kappag$) involving the gluon~\cite{Hosch_Whishant}. To date, the best constraints on these processes are from the DESY $ep$ Collider (HERA): $\kappag/\Lambda < 0.4~{\rm TeV^{-1}}$, at $95\%$ C.L.~\cite{hera_gluon_fcnc}, where $\Lambda$ is the new physics cut-off scale. 

In this analysis, we consider events where the top quark decays into a $b$~quark and a $W$ boson, and the latter subsequently decays leptonically ($W \rightarrow \ell \nu$, where $\ell~=~e,~\mu~{\rm or}~ \tau$, with the $\tau$ decaying to either an electron or a muon, and two neutrinos). This gives rise to an event with a charged lepton of high transverse momentum ($p_T$), significant missing transverse energy ($\MET$) from the neutrinos, and at least two jets, one that is a $b$-quark jet 
(from the top quark decay), and the other from a $c$ quark, $u$ quark, or a gluon. Displaced secondary vertices are used to identify $b$ jets~\cite{Abazov:2005zz}. The largest physics backgrounds to these events are from SM production of $W$+jets and $\ttbar$, along with smaller contributions from SM production of single top quarks ($tb$ and $tqb$) and dibosons ($WW$ and $WZ$). An additional source of background is from multijet events in which a jet is incorrectly identified as an electron or in which a muon from a heavy flavor decay appears isolated. 

The \dzero detector is described elsewhere~\cite{D0detector}. We use the same dataset, basic event selections and background modeling as in our SM single top quark search~\cite{Abazov:2005zz}; however, since the FCNC signal processes have only one $b$ quark in the final state, we consider here events with only one $b$-tagged jet. In addition, we include here the SM single top quark processes ($tb$ and $tqb$) in the background model. The data were recorded between August 2002 and March 2004 with a total integrated luminosity of $230 \pm 15 ~\rm pb^{-1}$~\cite{Edwards:2004jz} and were collected using a trigger that required a reconstructed jet and an electromagnetic energy cluster in the electron channel, or a jet and a muon in the muon channel.

We model the FCNC signal kinematics using a parton-level leading order (LO) matrix element event generator {\sc CompHEP}~\cite{comphepref}.  We consider the following four sub-processes:
\begin{equation}
\label{eq:subprocesses}
cq\!\!\!^{\footnotesize^{(\!-\!)}} \rightarrow tq\!\!\!^{\footnotesize^{(\!-\!)}}, ~~cg \rightarrow tg, ~~q\bar{q} \rightarrow t\bar{c}, ~~gg  \rightarrow t\bar{c},
\end{equation}
and also those that replace the $c$ quark with a $u$ quark and the charge conjugates. The identity of the associated final state jet depends upon the initial state of the system. Decays of the top quark and $W$ boson are done in {\sc CompHEP} to take into account all spin-dependent effects. The effects of FCNC couplings are parameterized in a model-independent way via an effective Lagrangian~\cite{Hosch_Whishant} that is a linear function of the factor $\kappag/\Lambda$. The production cross section of single top quarks thus depends quadratically on $\kappag/\Lambda$, and for certain values of $\kappag/\Lambda$ can be significantly larger than that in the SM, as shown in Table~\ref{tab:theory_xsec}. The cross sections are evaluated at a top quark mass of $m_t = 175~\rm GeV$, with the factorization and renormalization scales set to $Q^2 = m_t^2$. The LO cross sections are scaled to next-to-leading (NLO) order by a $K$-factor (NLO/LO cross section ratio) of 1.6~\cite{Liu}. 
\begin{table}[!h!tbp]
\caption{The production cross sections $\sigma(t)$ of single top quarks through a gluon exchange in $p\bar{p}$ collisions at $\sqrt{s}~=~1.96 \rm ~TeV$ for different $\kappa_g/\Lambda$ values, as obtained from {\sc CompHEP} and scaled to NLO by a $K$-factor of 1.6.}
\label{tab:theory_xsec}
\begin{center}
\begin{tabular}{lr@{.}lr@{.}l}\hline\hline
$\kappa_g/\Lambda$ [$\rm TeV^{-1}$] & \multicolumn{4}{c}{$\sigma(t)$ [pb]} \\
& \multicolumn{2}{c}{$tcg$} & \multicolumn{2}{c}{$tug$} \\
& \multicolumn{2}{c}{($\kappau$ = 0)} & \multicolumn{2}{c}{($\kappac$ = 0)} \\\hline
0.01     & 0& 05 & 0&88 \\ 
0.03     & 0&45   & 7&92 \\  
0.07     & 2&40   & 42&61 \\  
0.11     & 5&86   & 104&78 \\ \hline\hline
\end{tabular}
\end{center}
\vspace*{-0.5cm}
\end{table}

The effect of FCNC couplings on the top quark decay is negligible for $\kappag/\Lambda \lsim 0.2~{\rm TeV^{-1}}$~\cite{Hosch_Whishant}. In this range of $\kappag/\Lambda$, it is therefore safe to assume that the top quark decays into a $W$ boson and a $b$ quark with a branching fraction close to unity, as in the standard model, and hence, the cross section $\sigma(t)$ multiplied by the branching fraction for the process $t \rightarrow Wb \rightarrow \ell\nu b$ would also depend quadratically on $\kappag/\Lambda$. We may therefore model the shapes of the signal kinematic variables at any one value of $\kappag/\Lambda$  and scale the distributions appropriately to obtain them at any other value of the coupling. We choose that value of $\kappag/\Lambda$ to be $0.03~{\rm TeV^{-1}}$ in {\sc CompHEP} and generate two sets of signal events: one for the $tcg$ process only, in which $\kappau$ is set to zero, and the other for the $tug$ process only, in which $\kappac$  is set to zero.

The parton-level samples from {\sc CompHEP} are processed with {\sc pythia}~\cite{pythiaref} for fragmentation, hadronization, and modeling of the underlying event, using the {\sc cteq5l}~\cite{cteqref} parton distribution functions. We use {\sc tauola}~\cite{Jadach:1990mz} for the tau lepton decays and {\sc evtgen}~\cite{Lange:2001uf} for the $b$-hadron decays. The generated events are processed through a {\sc geant}-based~\cite{geantref} simulation of the \dzero detector, and normalized to the NLO cross sections for $\kappag/\Lambda = 0.03~{\rm TeV^{-1}}$. For the backgrounds, the Monte Carlo (MC) simulated samples are generated and normalized as described in Ref.~\cite{Abazov:2005zz}. 


The event selections~\cite{Abazov:2005zz} applied to the simulated signals and backgrounds and to the \dzero data are summarized in Table~\ref{tab:selections}. The resulting numbers of events from all samples, along with their systematic uncertainties described later, are shown in Table~\ref{tab:yields}. We find that the observed numbers of events agree with the predicted numbers for the SM backgrounds within uncertainties in both the electron and muon channels, and that the FCNC signals are a tiny fraction. We therefore construct multivariate discriminants using neural networks to separate the expected signal from the background and enhance the sensitivity.  
\begin{table}[!h!tbp]
\caption{Summary of event selections.} 
\begin{tabular}{l|cc}\hline\hline
&Electron channel & Muon channel \\ \hline
&& \\
Lepton &$E_T>$ 15 GeV & $p_T>$ 15 GeV   \\
&$|\eta|<$ 1.1 & $|\eta|<$ 2.0 \\\hline
&&\\
$\MET$ & \multicolumn{2}{c}{15 $<$ $\MET <$ 200 GeV} \\\hline
&&\\
Jets& \multicolumn{2}{c}{2, 3 or 4 jets, $E_T>$ 15 GeV, $|\eta|<$ 3.4}\\
&\multicolumn{2}{c}{$E_T$(jet1) $>$ 25 GeV, $|\eta({\rm jet1})|<$ 2.5}\\\hline
&&\\
$b$ jets&\multicolumn{2}{c}{exactly one tagged jet}  \\
\hline\hline
\end{tabular}
\label{tab:selections}
\end{table}
%
\begin{table}[!h!tbp]
\begin{center}
\caption{Estimates of the FCNC signal and background yields, and the number
of observed events in data after all selections, for the electron and muon channels. The signal yields are evaluated at $\kappag/\Lambda = 0.03~{\rm TeV^{-1}}$. The yields for $t\bar{t}$ include both lepton+jets and dilepton final states, and those from $W$+jets also include the diboson backgrounds.} 
\begin{tabular}{lr@{$\,\pm \,$}lr@{$\,\pm \,$}l}\hline\hline
Source           & \multicolumn{2}{c}{Electron channel} &
\multicolumn{2}{c}{Muon channel} \\ \hline
$tcg$ &   0.6 & 0.2 & 0.6 & 0.2  \\ 
$tug$ &   8.4 & 2.1 & 9.8 & 2.7  \\ \hline
SM single top ($tb$+$tqb$) &   6.4 & 1.4 & 6.1 & 1.4  \\ 
$t\bar{t}$    &   31.8 & 6.9 & 31.4 & 7.0  \\ 
$W$+jets         & 84.6 & 10.2 & 76.8 & 8.5  \\
Multijets         &  13.7 & 4.3 & 17.2 & 1.5  \\ \hline
Total SM background & 136.5 & 13.4 & 131.5 & 12.7 \\\hline
Observed no. of events  & \multicolumn{2}{c}{\hspace*{-0.8cm} 134}  &  \multicolumn{2}{c}{\hspace*{-0.25cm} 118} \\\hline\hline
\end{tabular}
\label{tab:yields}
\end{center}
\vspace*{-0.5cm}
\end{table}
%

\begin{figure}[!h!tbp]
\begin{center}
\includegraphics[width=0.3\textwidth]
{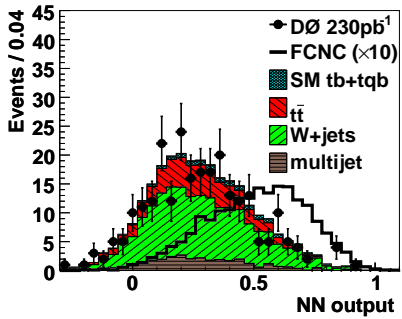}
\end{center}
\vspace{-0.7cm}
\caption{Neural network output distributions of summed background samples and \dzero data, for the combined electron and muon channels. Also shown is the FCNC signal distribution with the $tcg$ and $tug$ processes evaluated at $\kappag/\Lambda = 0.03~{\rm TeV^{-1}}$ and summed (color online).}
\label{fig:nnout}
\end{figure}

We use MLPfit implementation~\cite{mlpfit} of neural networks with ten input variables, one hidden layer, and one output layer. The input variables represent individual object kinematics, global event kinematics, and angular correlations, and are listed in~Table~\ref{tab:inputvars}. For training of the networks, we consider the sum of $tcg$ and $tug$ processes as signal, since the final states for these two processes are indistinguishable in this analysis, and the sum of all SM processes as background. The processes in each sum are weighted according to their relative proportions in the 230 $\rm pb^{-1}$ of data. The resulting networks separate the FCNC signals not only from the dominant backgrounds ($W$+jets and $t\bar{t}$) but also from the SM single top quark processes. Output distributions for the combined electron and muon channels are shown in Fig.~\ref{fig:nnout} for the summed background samples and \dzero data. Also shown is the FCNC signal with the $tcg$ and $tug$ processes evaluated at $\kappag/\Lambda = 0.03~{\rm TeV^{-1}}$ and summed. Since the observed spectrum agrees well with the predicted SM background, we set upper limits on the FCNC coupling parameters $\kappacLambda$ and $\kappauLambda$.
\begin{table*}[!h!tbp]
\caption[tab:inputvars]{Input variables used in the neural network analysis.}
\begin{center}
\begin{footnotesize}
\begin{tabular}{ll} \hline \hline
 $p_T({\rm jet1})$ & Transverse momentum of the leading jet\\
$p_T({\rm jet1}_{\rm tagged})$ & Transverse momentum of the $b$-tagged jet \\
$\eta({\rm lepton})$ & Pseudorapidity~\cite{pseudorapidity_endnote} of the lepton\\
{\MET} & Missing transverse energy\\ 
$p_T({\rm jet1},{\rm jet2})$ & Transverse momentum of the two leading jets \\
$H_T({\rm jet1},{\rm jet2})$ & Scalar sum of the transverse momenta of the two leading jets \\
$p_T(W)$ & Transverse momentum of the reconstructed $W$ boson \\
$M(W,{\rm jet1}_{\rm tagged})$ & Invariant mass of the reconstructed top quark using the $W$ boson~\cite{pzneutrino_endnote} and the $b$-tagged jet \\
$M({\rm alljets})$ & Invariant mass of all jets \\
$\cos({\rm jet1}, {\rm lepton})_{\rm lab}$ & Cosine of the angle between the leading jet and lepton in the laboratory frame of reference \\
\hline \hline
\end{tabular}
\end{footnotesize}
\label{tab:inputvars}
\end{center}
\vspace*{-0.5cm}
\end{table*}
%

To estimate systematic uncertainties, we consider two classes of effects: those that alter the overall normalization of the distributions and those that also change their shapes. The dominant normalization effects are from lepton identification (4\%), integrated luminosity measurement (6.5\%), and cross section estimates. The uncertainties on the cross sections vary from 9\% for diboson production to 16\% for SM single top quark production and 18\% for the $\ttbar$ samples~\cite{Kidonakis:2003qe}. The latter two include the uncertainty due to the top quark mass. For the FCNC signal, we factor out the parameter $(\kappag/\Lambda)^2$ from the cross section, and assume an uncertainty of $15\%$ on the remaining quantity based on a discussion in Ref.~\cite{Liu} on how the theoretical predictions depend on the particular choice of the factorization scale. The $W$+jets and multijets samples have an overall uncertainty of  $4\%$ from their normalization to data~\cite{Abazov:2005zz}. This includes an uncertainty of $25\%$ on the heavy flavor fraction of the $W$+jets sample.

The shape effects are modeled by shifting the uncertainty components one-by-one by plus or minus one standard deviation with respect to their nominal values, for each sample, and propagating the changes to the kinematics of the different objects (electrons, muons, jets, and $\MET$) before making any event selections. The resulting uncertainties are as follows: (i) (1--16)\% due to jet energy scale, (ii) (2--8)\% from trigger modeling, (iii) (1--5)\% due to jet energy resolution, (iv) (1--9)\% due to jet identification, and (v) (5--13)\% from $b$-tag modeling. Although the $W$+jets MC yield is normalized to data, it is also affected by the uncertainty from the $b$-tag modeling since the normalization is done before $b$-tag parametrization, and we take this into account.

We use a Bayesian approach to set upper limits~\cite{IainTM2000} on the FCNC coupling parameters. Given $N$ observed events, we define the Bayesian posterior probability density in a two-dimensional plane of $(\kappacLambda)^2$ and $(\kappauLambda)^2$ as:
\begin{widetext}
\begin{equation}
\label{eq:posterior}
{p}([\kappacLambda]^2, [\kappauLambda]^2~ |~ N) ~\propto ~\int\int\int~{L}(N~ |~ n) ~{p_1}(f_c, f_u, b) ~{p_2}([\kappacLambda]^2)~ {p_3}([\kappauLambda]^2)~ {\rm d}{f_c}{\rm d}{f_u}{\rm d}{b}, 
\end{equation}
\end{widetext}
where $L$ is a Poisson likelihood with mean $n$, and $p_i ~{\rm (} i = 1, 2, 3{\rm )}$ are the prior probability densities of the respective parameters. The likelihood $L$ is a product of the likelihoods over all bins of the neural network output distributions, $n$ is the predicted number of events, equal to the sum of signal ($s$) and background ($b$) yields:
\begin{eqnarray}
\label{eq:predicted}
n &=& s~+~b \\\nonumber
&= & f_c ~\times ~(\kappacLambda)^2 ~+~ f_u ~\times ~ (\kappauLambda)^2 ~+~ b, 
\end{eqnarray}
where the constants $f_c$ and $f_u$ are determined from the simulated signal samples at $\kappag/\Lambda~=~0.03~{\rm TeV^{-1}}$. The prior probability density, $p_1$,  is a multivariate Gaussian, with the mean and standard deviation defined by the estimated yields and their uncertainties, to take into account correlations among the different samples and bins. Since the signal cross sections depend quadratically on $\kappag/\Lambda$, for $p_2$ and $p_3$ we choose priors flat in $(\kappacLambda)^2$ and $(\kappauLambda)^2$ respectively, which imply priors flat in the corresponding cross sections.

\begin{figure}[!h!tbp]
\begin{center}
\includegraphics[width=0.30\textwidth]
{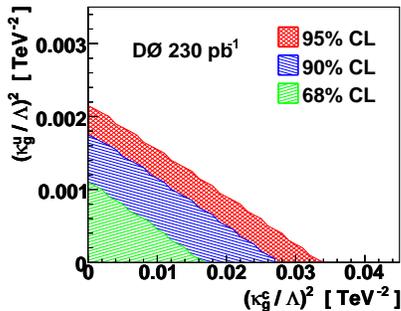}
\end{center}
\vspace{-0.7cm}
\caption{Exclusion contours at various levels of confidence using 230 $\rm pb^{-1}$ of \dzero data in both the electron and muon channels (color online).}
\label{fig:quad}
\end{figure}

From the two-dimensional posterior probability density, exclusion contours at different levels of confidence ($k$) are defined as contours of equal probability that enclose a volume $k$ around the peak of the posterior probability density. These contours are shown in Fig.~\ref{fig:quad}, using data from both the electron and muon channels and including all systematic uncertainties with correlations. The one-dimensional posterior probability density over any dimension is obtained by integrating the two-dimensional posterior over the other dimension. The resulting limits, translated to $\kappag/\Lambda$, using data (observed limits) as well as the expected limits for which the observed count is set to the predicted background yield in any bin, are summarized in Table~\ref{tab:quad}.

To conclude, we analyzed 230 $\rm pb^{-1}$ of lepton+jets data collected at \dzero from $\ppbar$ collisions at a center of mass energy of 1.96 TeV, and searched for the presence of non-SM production of single top quarks. We found no deviation from SM predictions, and therefore set limits on the anomalous coupling parameters, $\kappacLambda$ and $\kappauLambda$, using multivariate neural network discriminants. The 95\% C.L. observed (expected) limits are 0.15 (0.16) $\rm TeV^{-1}$ on  $\kappacLambda$, and 0.037 (0.041) $\rm TeV^{-1}$ on $\kappauLambda$. These are the first limits from hadron colliders on FCNC couplings of a gluon to the top quark and a charm or up quark, and a factor 3--11 better than those from HERA. 
\begin{table}[!h!tbp]
\caption[quadratic-limits]{Upper limits on $\kappacLambda$ and $\kappauLambda$, at 95\% C.L.}
\label{tab:quad}
\begin{center}
\begin{ruledtabular}
\begin{tabular}{l|cc}
&\multicolumn{2}{c}{Observed (expected) limits [$\rm TeV^{-1}$]} \\
&$\kappacLambda$ & $\kappauLambda$\\
\hline
Electron channel   & 0.16  (0.19)&  0.046 (0.052)      \\ 
Muon channel       & 0.21  (0.21)  & 0.049 (0.050)     \\ 
Combined             & \textbf{0.15}  \textbf{(0.16)}&  \textbf{0.037}  \textbf{(0.041)}
\end{tabular}
\end{ruledtabular}
\end{center}
\vspace*{-0.5cm}
\end{table}

We are grateful to Tim Tait for discussions related to this search.
%
We thank the staffs at Fermilab and collaborating institutions, 
and acknowledge support from the 
DOE and NSF (USA);
CEA and CNRS/IN2P3 (France);
FASI, Rosatom and RFBR (Russia);
CAPES, CNPq, FAPERJ, FAPESP and FUNDUNESP (Brazil);
DAE and DST (India);
Colciencias (Colombia);
CONACyT (Mexico);
KRF and KOSEF (Korea);
CONICET and UBACyT (Argentina);
FOM (The Netherlands);
PPARC (United Kingdom);
MSMT (Czech Republic);
CRC Program, CFI, NSERC and WestGrid Project (Canada);
BMBF and DFG (Germany);
SFI (Ireland);
The Swedish Research Council (Sweden);
Research Corporation;
Alexander von Humboldt Foundation;
and the Marie Curie Program.
%


\end{document}

%% file: list_of_authors_r2.tex
%
\author{                                                                      
V.M.~Abazov,$^{35}$                                                           
B.~Abbott,$^{75}$                                                             
M.~Abolins,$^{65}$                                                            
B.S.~Acharya,$^{28}$                                                          
M.~Adams,$^{51}$                                                              
T.~Adams,$^{49}$                                                              
E.~Aguilo,$^{5}$                                                              
S.H.~Ahn,$^{30}$                                                              
M.~Ahsan,$^{59}$                                                              
G.D.~Alexeev,$^{35}$
% ...
\\                                                                            
\vskip 0.30cm                                                                 
\centerline{(D\O\ Collaboration)}                                             
\vskip 0.30cm                                                                 
}                                                                             
\affiliation{                                                                 
\centerline{$^{1}$Universidad de Buenos Aires, Buenos Aires, Argentina}       
\centerline{$^{2}$LAFEX, Centro Brasileiro de Pesquisas F{\'\i}sicas,         
                  Rio de Janeiro, Brazil}                                     
\centerline{$^{3}$Universidade do Estado do Rio de Janeiro,                   
                  Rio de Janeiro, Brazil}                                     
\centerline{$^{4}$Instituto de F\'{\i}sica Te\'orica, Universidade            
                  Estadual Paulista, S\~ao Paulo, Brazil}                     
\centerline{$^{5}$University of Alberta, Edmonton, Alberta, Canada,           
                  Simon Fraser University, Burnaby, British Columbia, Canada,}
\centerline{York University, Toronto, Ontario, Canada, and                    
                  McGill University, Montreal, Quebec, Canada}                
\centerline{$^{6}$University of Science and Technology of China, Hefei,       
                  People's Republic of China}                                 
\centerline{$^{7}$Universidad de los Andes, Bogot\'{a}, Colombia}             
\centerline{$^{8}$Center for Particle Physics, Charles University,            
                  Prague, Czech Republic}                                     
\centerline{$^{9}$Czech Technical University, Prague, Czech Republic}         
\centerline{$^{10}$Center for Particle Physics, Institute of Physics,         
                   Academy of Sciences of the Czech Republic,                 
                   Prague, Czech Republic}                                    
\centerline{$^{11}$Universidad San Francisco de Quito, Quito, Ecuador}        
\centerline{$^{12}$Laboratoire de Physique Corpusculaire, IN2P3-CNRS,         
                   Universit\'e Blaise Pascal, Clermont-Ferrand, France}      
\centerline{$^{13}$Laboratoire de Physique Subatomique et de Cosmologie,      
                   IN2P3-CNRS, Universite de Grenoble 1, Grenoble, France}    
\centerline{$^{14}$CPPM, IN2P3-CNRS, Universit\'e de la M\'editerran\'ee,     
                   Marseille, France}                                         
\centerline{$^{15}$Laboratoire de l'Acc\'el\'erateur Lin\'eaire,              
                   IN2P3-CNRS et Universit\'e Paris-Sud, Orsay, France}       
\centerline{$^{16}$LPNHE, IN2P3-CNRS, Universit\'es Paris VI and VII,         
                   Paris, France}                                             
\centerline{$^{17}$DAPNIA/Service de Physique des Particules, CEA, Saclay,    
                   France}                                                    
\centerline{$^{18}$IPHC, IN2P3-CNRS, Universit\'e Louis Pasteur, Strasbourg,  
                   France, and Universit\'e de Haute Alsace,                  
                   Mulhouse, France}                                          
\centerline{$^{19}$IPNL, Universit\'e Lyon 1, CNRS/IN2P3, Villeurbanne, France
                   and Universit\'e de Lyon, Lyon, France}                    
\centerline{$^{20}$III. Physikalisches Institut A, RWTH Aachen,               
                   Aachen, Germany}                                           
\centerline{$^{21}$Physikalisches Institut, Universit{\"a}t Bonn,             
                   Bonn, Germany}                                             
\centerline{$^{22}$Physikalisches Institut, Universit{\"a}t Freiburg,         
                   Freiburg, Germany}                                         
\centerline{$^{23}$Institut f{\"u}r Physik, Universit{\"a}t Mainz,            
                   Mainz, Germany}                                            
\centerline{$^{24}$Ludwig-Maximilians-Universit{\"a}t M{\"u}nchen,            
                   M{\"u}nchen, Germany}                                      
\centerline{$^{25}$Fachbereich Physik, University of Wuppertal,               
                   Wuppertal, Germany}                                        
\centerline{$^{26}$Panjab University, Chandigarh, India}                      
\centerline{$^{27}$Delhi University, Delhi, India}                            
\centerline{$^{28}$Tata Institute of Fundamental Research, Mumbai, India}     
\centerline{$^{29}$University College Dublin, Dublin, Ireland}                
\centerline{$^{30}$Korea Detector Laboratory, Korea University,               
                   Seoul, Korea}                                              
\centerline{$^{31}$SungKyunKwan University, Suwon, Korea}                     
\centerline{$^{32}$CINVESTAV, Mexico City, Mexico}                            
\centerline{$^{33}$FOM-Institute NIKHEF and University of                     
                   Amsterdam/NIKHEF, Amsterdam, The Netherlands}              
\centerline{$^{34}$Radboud University Nijmegen/NIKHEF, Nijmegen, The          
                  Netherlands}                                                
\centerline{$^{35}$Joint Institute for Nuclear Research, Dubna, Russia}       
\centerline{$^{36}$Institute for Theoretical and Experimental Physics,        
                   Moscow, Russia}                                            
\centerline{$^{37}$Moscow State University, Moscow, Russia}                   
\centerline{$^{38}$Institute for High Energy Physics, Protvino, Russia}       
\centerline{$^{39}$Petersburg Nuclear Physics Institute,                      
                   St. Petersburg, Russia}                                    
\centerline{$^{40}$Lund University, Lund, Sweden, Royal Institute of          
                   Technology and Stockholm University, Stockholm,            
                   Sweden, and}                                               
\centerline{Uppsala University, Uppsala, Sweden}                              
\centerline{$^{41}$Physik Institut der Universit{\"a}t Z{\"u}rich,            
                   Z{\"u}rich, Switzerland}                                   
\centerline{$^{42}$Lancaster University, Lancaster, United Kingdom}           
\centerline{$^{43}$Imperial College, London, United Kingdom}                  
\centerline{$^{44}$University of Manchester, Manchester, United Kingdom}      
\centerline{$^{45}$University of Arizona, Tucson, Arizona 85721, USA}         
\centerline{$^{46}$Lawrence Berkeley National Laboratory and University of    
                   California, Berkeley, California 94720, USA}               
\centerline{$^{47}$California State University, Fresno, California 93740, USA}
\centerline{$^{48}$University of California, Riverside, California 92521, USA}
\centerline{$^{49}$Florida State University, Tallahassee, Florida 32306, USA} 
\centerline{$^{50}$Fermi National Accelerator Laboratory,                     
            Batavia, Illinois 60510, USA}                                     
\centerline{$^{51}$University of Illinois at Chicago,                         
            Chicago, Illinois 60607, USA}                                     
\centerline{$^{52}$Northern Illinois University, DeKalb, Illinois 60115, USA} 
\centerline{$^{53}$Northwestern University, Evanston, Illinois 60208, USA}    
\centerline{$^{54}$Indiana University, Bloomington, Indiana 47405, USA}       
\centerline{$^{55}$University of Notre Dame, Notre Dame, Indiana 46556, USA}  
\centerline{$^{56}$Purdue University Calumet, Hammond, Indiana 46323, USA}    
\centerline{$^{57}$Iowa State University, Ames, Iowa 50011, USA}              
\centerline{$^{58}$University of Kansas, Lawrence, Kansas 66045, USA}         
\centerline{$^{59}$Kansas State University, Manhattan, Kansas 66506, USA}     
\centerline{$^{60}$Louisiana Tech University, Ruston, Louisiana 71272, USA}   
\centerline{$^{61}$University of Maryland, College Park, Maryland 20742, USA} 
\centerline{$^{62}$Boston University, Boston, Massachusetts 02215, USA}       
\centerline{$^{63}$Northeastern University, Boston, Massachusetts 02115, USA} 
\centerline{$^{64}$University of Michigan, Ann Arbor, Michigan 48109, USA}    
\centerline{$^{65}$Michigan State University,                                 
            East Lansing, Michigan 48824, USA}                                
\centerline{$^{66}$University of Mississippi,                                 
            University, Mississippi 38677, USA}                               
\centerline{$^{67}$University of Nebraska, Lincoln, Nebraska 68588, USA}      
\centerline{$^{68}$Princeton University, Princeton, New Jersey 08544, USA}    
\centerline{$^{69}$State University of New York, Buffalo, New York 14260, USA}
\centerline{$^{70}$Columbia University, New York, New York 10027, USA}        
\centerline{$^{71}$University of Rochester, Rochester, New York 14627, USA}   
\centerline{$^{72}$State University of New York,                              
            Stony Brook, New York 11794, USA}                                 
\centerline{$^{73}$Brookhaven National Laboratory, Upton, New York 11973, USA}
\centerline{$^{74}$Langston University, Langston, Oklahoma 73050, USA}        
\centerline{$^{75}$University of Oklahoma, Norman, Oklahoma 73019, USA}       
\centerline{$^{76}$Oklahoma State University, Stillwater, Oklahoma 74078, USA}
\centerline{$^{77}$Brown University, Providence, Rhode Island 02912, USA}     
\centerline{$^{78}$University of Texas, Arlington, Texas 76019, USA}          
\centerline{$^{79}$Southern Methodist University, Dallas, Texas 75275, USA}   
\centerline{$^{80}$Rice University, Houston, Texas 77005, USA}                
\centerline{$^{81}$University of Virginia, Charlottesville,                   
            Virginia 22901, USA}                                              
\centerline{$^{82}$University of Washington, Seattle, Washington 98195, USA}  
}                                                                             

%% file: fcnc_prl.bbl
\begin{thebibliography}{99}
\bibitem{pseudorapidity_endnote}
Pseudorapidity is defined as $\eta = - \ln(\tan \frac{\theta}{2})$, 
where $\theta$ is the polar angle with respect to the beam axis,
with the origin at the primary vertex.
\bibitem{pzneutrino_endnote}
The longitudinal component of the momentum of the neutrino is obtained using a SM $W$ boson mass constraint, choosing the smaller of the two possible solutions.
\end{thebibliography}
